\documentclass{article}%
\usepackage[english,russian]{babel}
\usepackage[cp1251]{inputenc}
\usepackage{amssymb}
\usepackage{amsmath}

\numberwithin{equation}{section}

\title{\textbf{ОБОБЩЕННОЕ ВЗАИМОДЕЙСТВИЕ В 
МУЛЬТИГРАВИТАЦИИ\thanks{Принято в ТМФ.}}%
}
\author{\textbf{С. А. Дуплий,}\\ физико-технический факультет,\\
Харьковский национальный университет им. В.Н. Каразина,\\
пл. Свободы, 4, г. Харьков, 61022, Украина,\\ 
\textbf{А. Т. Котвицкий,}\\ физический факультет,\\
Харьковский национальный университет им. В.Н. Каразина,\\
пл. Свободы, 4, г. Харьков, 61022, Украина}

\begin{document}
\maketitle
\begin{abstract}
Рассматривается общий подход к описанию взимодействия моделей
мультигравитации в $D$-мерном пространстве-времени.
Приведены различные возможности для обобщения инвариантного
объема. Далее конструируется наиболее общий вид потенциала
взаимодействия, который в случае бигравитации
переходит в модель типа Паули-Фирца.
Подробный анализ данной модели, проведенный в формализме
3+1 разложения, а также требование отсутствия духов, приводят
 к тому,
что данная бигравитационная модель в пределе слабого поля
полностью эквивалентна модели Паули-Фирца.
Таким образом, на конкретном примере показано, что введение взаимодействия между метриками
 эквивалентно введению массы у гравитона.
\end{abstract}

\par\noindent {\bf КЛЮЧЕВЫЕ
СЛОВА}: мультигравитация, бигравитация, массивная гравитация, инвариантный объем,
потенциал взаимодействия, модель Паули-Фирца

\section{ВВЕДЕНИЕ}

Мультигравитация,
наряду с
конформной
гравитацией
\cite{mann06} и
скалярными
теориями \cite{goe04},
является
одним из
возможных
расширений
общей теории
относительности
\cite{weinberg72,wald}. В первых
работах
частный
случай
мультигравитации
(бигравитация)
называлась
\textquotedblleft\textit{f-g} theory\textquotedblright\  или
\textquotedblleft strong gravity\textquotedblright%
\ \cite{ish/sal/str,aic/man/urb,aic}). В
дальнейшем
эта
конструкция
успешно
применялась
в квантовой
гравитации и
бранах
\cite{kog/ros,kog/mou/pap/ros,kog/mou/pap},
теориях с
дискретными
размерностями
\cite{def/mou1,def/mou2}, теории
перенормировок
\cite{gar1}, массивной
гравитации
\cite{bla}, а также в
объяснении
таких
экспериментальных
фактов, как
темная
энергия и
материя,
\cite{han,gri/pav,dub/tin/tka},
ускоренное
расширение
вселенной
\cite{dam/kog/pap,def/dva/gab}.
Поэтому
важным
является
рассмотрение
нелинейных
формулировок
мультигравитации
(для
бигравитации
это
рассматривалось
в \cite{dam/kog}).

С другой
стороны, в
теории
массивной
гравитации
прогресс был
связан с
работами \cite{rha/gab10},
где было
продемонстрировано
расширение
массового
слагаемого
Паули-Фирца
для
линеаризованой
теории
гравитации и
показано,
что в такой
модели нет
духовых мод
\cite{bou/des72}. Далее
теория была
расширена на
общую
дополнительную
метрику \cite{koy/niz/tas}.
Основные
свойства
подобных
теорий были
рассмотрены
в работах
\cite{rha/gab/tol,cha/muk}, а в \cite{has/ros12}
было
проведено
доказательство
отсутствия
духовых
слагаемых в
нелинейных
моделях. В
теориях
гравитации с
ненулевой
массой
существует
особенность,
которая
проявляется
в том, что при
стремлении
массы
гравитона к
нулю теория
не переходит
в ОТО \cite{zakh70,dam/vel}.
Механизм
Вайнштейна
\cite{vai72} позволяет
избежать
такой
неоднородности
в
пространстве
параметров
\cite{koy/niz/tas,bab/def/zio}, и, кроме
того, такая
неоднородность
может
устраняться
в случае
неплоской
фоновой
метрики \cite{rub/tin,hin12}.

В данной
работе мы
рассматриваем
общий подход
к описанию
взимодействия
моделей
мультигравитации
в $D$ мерном
пространстве-времени.
В первой
части
изучаются
различные
возможности
для
обобщения
инвариантного
объема $d\Omega_{int}^{\left(  N\right)
}$, на который
накладываются
ограничения
состоящие в
том, что $d\Omega_{int}^{\left(  N\right)
}$ должен быть
скаляром, в
пределе
совпадения
всех метрик
инвариантный
объем должен
переходить в
стандартный
$\sqrt{g}d^{D}x.$ Также
функция $d\Omega_{int}^{\left(
N\right)  }$ должна
быть
монотонной и
однородной
по всем
метрикам $g_{i}$. В
следующем
разделе мы
конструируем
наиболее
общий вид
потенциала
взаимодействия.
И
показываем,
что в самом
простом
случае двух
метрик
(бигравитации)
он переходит
в модель
типа
Паули-Фирца.
Подробный
анализ
данной
модели,
проведенный
в формализме
3+1 разложении
и требование
отсутствия
духов
приводит к
тому, что
данная
бигравитационная
модель в
пределе
слабого поля
полностью
эквивалентна
модели
Паули-Фирца.
Фактически
это
означает,
что введение
взаимодействия
между
тензорными
полями $g_{\mu\nu}^{(1)}$ и
$g_{\mu\nu}^{(2)}$ может
быть
эквивалентно
введению
массы у гравитона.

В приложении
мы приводим
новый способ
вычисления
$\sqrt{g}$ для случая
малых
поправок,
который
является
пригодным
для любой
фоновой
метрики. В
случае
плоского
фонового
пространства-времени
Минковского
получается
стандартное выражение.

\section{МУЛЬТИГРАВИТАЦИЯ
И ОБОБЩЕНИЕ
ИНВАРИАНТНОГО
ОБЪЕМА
ВЗАИМОДЕЙСТВИЯ
\label{sec-inv}}

Рассмотрим
совокупность
$N$ различных
вселенных,
каждая из
которых
описывается
метрикой $\mathsf{g}_{\mu\nu
}^{\left(  i\right)  }$, где $i=1,\ldots N$. В
$D$-мерном
пространстве-времени
мы
используем
сигнатуру $\left(
+,\overset{D-1}{\overbrace{-,\ldots-}}\right)  $. Для
$i$-той
вселенной
действие
запишем в
виде%
\begin{equation}
S_{G\left(  i\right)  }=\int d\Omega^{\left(  i\right)  }\left[
L_{gr}^{\left(  i\right)  }(\mathsf{g}^{\left(  i\right)  })+L_{mat}\left(
\mathsf{g}^{\left(  i\right)  },\Phi^{\left(  i\right)  }\right)  \right]  ,
\label{s}%
\end{equation}
где $d\Omega^{\left(  i\right)  }=d^{4}x\sqrt{g^{\left(
i\right)  }}$, $g^{\left(  i\right)  }=\left\vert \det\left(  \mathsf{g}%
_{\mu\nu}^{\left(  i\right)  }\right)  \right\vert $ ---
инвариантный
объем, $g^{\left(  i\right)  }$ ---
скалярная
плотность
веса 2 и $\mathsf{g}_{\mu\nu}^{\left(  i\right)
}$ ---
метрический
тензор в $i$-той
вселенной,
$L_{gr}^{\left(  i\right)  }(\mathsf{g}^{\left(  i\right)  })$ ---
лагранжиан,
описывающий
гравитационное
поле, $L_{mat}\left(  \mathsf{g}^{\left(  i\right)
},\Phi^{\left(  i\right)  }\right)  $
описывает
взаимодействие
гравитации и
материальных
полей $\Phi^{\left(  i\right)  }$. 
При этом, интегрирование в (\ref{s}) ведется по общему 
 многообразию для $N$ вселенных.

В
предположении
\textquotedblleft слабо
связанных
миров\textquotedblright\ \cite{dam/kog} и
\textquotedblleft no-go\textquotedblright%
\  теоремы \cite{bou/dam/gua/hen}
общее
действие для
$N$
безмассовых
гравитонов
записывется
в виде суммы
чисто
гравитационных
действий (\ref{s})%
\begin{equation}
S_{0}=\sum_{i=1}^{N}S_{G\left(  i\right)  }.
\end{equation}

В
предположении,
что \textquotedblleft слабо
связанные
миры\textquotedblright%
\  взаимодействуют
только за
счет
гравитационных
полей,
полное
действие
мультигравитации
можно
записать в
виде суммы%

\begin{equation}
S_{full}=\sum_{i}^{N}S_{G\left(  i\right)  }+S_{int}, \label{sm}%
\end{equation}
где
последнее
слагаемое $S_{int}$
описывает
взаимодействие
вселенных.
Выбор этого
слагаемого
является
ключевым при
описании
моделей
мультигравитации
\cite{dup/kotv2007}.

В общем
случае $D$
измерений
для $N$%
-гравитации
$S_{int}$ можно
представить
как%
\begin{equation}
S_{int}=\int d^{D}xW(\mathsf{g}^{\left(  1\right)  },\ldots,\mathsf{g}%
^{\left(  N\right)  }),
\end{equation}
где $d^{D}x$ и $W(\mathsf{g}^{\left(  1\right)
},\ldots,\mathsf{g}^{\left(  N\right)  })\ $---
скалярные
плотности
противоположных
весов. По
аналогии со
стандартным
инвариантным
объемом $d\Omega=d^{4}x\sqrt{g}$
в общей
теории
относительности
\cite{weinberg72,wald},
представим
выражение $d^{D}%
xW(\mathsf{g}^{\left(  1\right)  },\ldots,\mathsf{g}^{\left(  N\right)  })$
как произведение%

\begin{equation}
d^{D}x\cdot f\left(  \sqrt{g_{1}},\ldots,\sqrt{g_{N}}\right)  \cdot
V(\mathsf{g}^{\left(  1\right)  },\ldots,\mathsf{g}^{\left(  N\right)  }),
\label{fv}%
\end{equation}
где $V(\mathsf{g}^{\left(  1\right)  },\ldots,\mathsf{g}%
^{\left(  N\right)  })$ ($\equiv V\left(  \mathsf{g}^{\left(  i\right)
}\right)  $) - скалярный
потенциал
взаимодействия
и $f\left(  \sqrt{g_{1}},\ldots,\sqrt{g_{N}}\right)  $ -
гладкая
положительная
функция с
весом $-1$
имеющая $N$
положительных
(действительных)
аргументов.
Введем
инвариантный
объем
взаимодействия%
\begin{equation}
d\Omega_{int}^{\left(  N\right)  }=d^{D}xf\left(  \sqrt{g_{1}},\ldots
,\sqrt{g_{N}}\right)  , \label{w}%
\end{equation}
который
должен быть
скаляром.
Кроме того, в
пределе
совпадения
\cite{dup/kotv2007} $\mathsf{g}_{\mu\nu}^{\left(  1\right)  }=\ldots
=\mathsf{g}_{\mu\nu}^{\left(  N\right)  }\equiv\mathsf{g}_{\mu\nu}$
инвариантный
объем
взаимодействия
должен
переходить в
стандартный
инвариантеый
объем $d\Omega_{int}^{\left(  N\right)
}\rightarrow d\Omega$. Для того
чтобы
удовлетворить
всем
вышеперечисленным
требованиям
функция $f\left(  \sqrt{g_{1}%
},\ldots,\sqrt{g_{N}}\right)  $ должна
быть: 1)
идемпотентная
в пределе
совпадения
$f\left(  \sqrt{g},\ldots,\sqrt{g}\right)  =\sqrt{g}$; 2)
монотонная; 3)
однородная
по всем
аргументам
$f\left(  t\sqrt{g_{1}},\ldots,t\sqrt{g_{N}}\right)  =t^{\alpha}f\left(
\sqrt{g_{1}},\ldots,\sqrt{g_{N}}\right)  $ (из
требования
идемпотентности
следует, что
$\alpha=1$); 4)
симметричная
по всем аргументам.

Из
требования
однородности
и
симметричности
функции $f\left(  \sqrt{g_{1}%
},\ldots,\sqrt{g_{N}}\right)  $
следует, что
инвариантный
объем
взаимодействия
может быть
представлен
как \cite{dup/kotv2007}%
\begin{equation}
d\Omega_{int}^{\left(  N\right)  }=d^{D}x\cdot f\left(  \sqrt{g_{1}}%
,...,\sqrt{g_{N}}\right)  =d^{D}x\sqrt[2N]{g_{1}...g_{N}}\cdot f\left(
y_{1}^{\left(  N\right)  },...,y_{N}^{\left(  N\right)  }\right)  ,
\end{equation}
где%
\begin{equation}
y_{1}^{\left(  N\right)  }=\sqrt[2N]{g_{1}^{N-1}g_{2}^{-1}g_{3}^{-1}%
...g_{N}^{-1}},\ldots\ \ \ \ y_{N}^{\left(  N\right)  }=\sqrt[2N]{g_{1}%
^{-1}g_{2}^{-1}...g_{N-1}^{-1}g_{N}^{N-1}}. \label{yy1}%
\end{equation}
Переменные
$y_{i}^{\left(  N\right)  }$
очевидно
удовлетворяют
тождеству%
\begin{equation}
y_{1}^{\left(  N\right)  }\cdot y_{2}^{\left(  N\right)  }\cdot\ldots\cdot
y_{N}^{\left(  N\right)  }=1, \label{y}%
\end{equation}
следовательно
функция $f$ в
действительности
является
функцией $N-1$
аргуметов и
инвариантный
объем
взаимодействия
можно
записать в
виде%
\begin{equation}
d\Omega_{int}^{\left(  N\right)  }=d^{4}x\cdot f\left(  \sqrt{g_{1}}%
,...,\sqrt{g_{N}}\right)  =d^{4}x\sqrt[2N]{g_{1}...g_{N}}\cdot\hat{f}\left(
y_{1}^{\left(  N\right)  },\ldots,y_{N-1}^{\left(  N\right)  }\right)  ,
\end{equation}
где $\hat{f}\left(  y_{1}^{\left(  N\right)  }%
,\ldots,y_{N-1}^{\left(  N\right)  }\right)  \overset{def}{=}f\left(
y_{1}^{\left(  N\right)  },\ldots,y_{N-1}^{\left(  N\right)  },\dfrac{1}%
{y_{1}^{\left(  N\right)  }\cdot y_{2}^{\left(  N\right)  }\ldots\cdot
y_{N-1}^{\left(  N\right)  }}\right)  $.
Заметим, что
в пределе
совпадения
$y_{i}^{\left(  N\right)  }=1$ и $f\left(  1,\ldots,1\right)  =1$.

Выберем
конкретный
вид
инвариантного
объема
взаимодействия
как
произвольную
сумму трех
средних:
среднее
арифметическое,
среднее
геометрическое,
среднее
гармоническое
с
произвольными
действительными
коэффициентами
$\alpha,\beta,\gamma$. Тогда
\begin{equation}
d\Omega_{int}^{\left(  N\right)  }=d^{D}x\cdot\sqrt[2N]{g_{1}...g_{N}}%
\cdot\frac{1}{\alpha+\beta+\gamma}\left[  \frac{\alpha}{N}\sum_{i=1}^{N}%
y_{i}^{\left(  N\right)  }+\beta+\gamma\frac{N}{\sum_{i=1}^{N}\dfrac{1}%
{y_{i}^{\left(  N\right)  }}}\right]  ,\label{dom}%
\end{equation}
где $\alpha+\beta+\gamma\neq0$.
Из соображений простоты, 
ограничимся этим естественным
выражением (\ref{dom})
инвариантного
объема
взаимодействия
в
мультигравитации.
Отметим, что
в \cite{dam/kog}, был
рассмотрен
частный
случай   (\ref{dom}) при $\alpha=\gamma=0$ и $\beta=1$ 
для бигравитации $N=2$.

\section{ОБОБЩЕННЫЙ
ПОТЕНЦИАЛ
ВЗАИМОДЕЙСТВИЯ}%

Рассмотрим
общий вид
взаимодействия
мультигравитации,
которое
описывается
скалярным
потенциалом
$V(\mathsf{g}^{\left(  1\right)  },\ldots,\mathsf{g}^{\left(  N\right)  })$
как функция
от $N$ метрик $\mathsf{g}%
_{\mu\nu}^{\left(  i\right)  }$ в $D$%
-мерном
пространстве-времени.
Группа
симметрии $N$
вселенных
является
прямым
произведением
групп
диффеоморфизмов
\cite{dam/kog}%
\begin{equation}
G_{full}=\mathrm{Diff}\left(  \varepsilon_{\mu}^{\left(  1\right)  }\right)
\times\mathrm{Diff}\left(  \varepsilon_{\mu}^{\left(  2\right)  }\right)
\times\ldots\times\mathrm{Diff}\left(  \varepsilon_{\mu}^{\left(  N\right)
}\right)  , \label{gf}%
\end{equation}
где каждый
диффеоморфизм
$\mathrm{Diff}\left(  \varepsilon_{\mu}^{\left(  i\right)  }\right)  $
действует на
метрику $\mathsf{g}_{\mu\nu
}^{\left(  i\right)  }$ вдоль
вектора $\varepsilon_{\mu}^{\left(
i\right)  }\left(  x\right)  $. По
известной
теореме \cite{bou/dam/gua/hen}
группа $G_{full}$
может быть
редуцирована
к
диагональной
подгруппе,
когда все
векторы
совпадают $\varepsilon
_{\mu}^{\left(  i\right)  }\left(  x\right)  =\varepsilon_{\mu}\left(
x\right)  $. Тогда
инфинитизимальные
преобразования
любой
метрики $\mathsf{g}_{\mu\nu
}^{\left(  i\right)  }$
определяются
производной
Ли%
\begin{equation}
\delta\mathsf{g}_{\mu\nu}^{\left(  i\right)  }=\mathcal{L}_{\varepsilon
}\mathsf{g}_{\mu\nu}^{\left(  i\right)  }=\varepsilon^{\rho}\partial_{\rho
}\mathsf{g}_{\mu\nu}^{\left(  i\right)  }+\mathsf{g}_{\mu\rho}^{\left(
i\right)  }\partial_{\nu}\varepsilon^{\rho}+\mathsf{g}_{\rho\nu}^{\left(
i\right)  }\partial_{\mu}\varepsilon^{\rho}. \label{dg}%
\end{equation}

Понятно, что
скалярный
потенциал
взаимодействия
должен быть
функцией от
скалярных
функций от
метрик $\mathsf{g}_{\mu\nu}^{\left(
i\right)  }$.
Естественным
выбором этих
скалярных
функций
могут
служить
инварианты
тензора с
одним
ковариантным
и одним
контравариантным
индексами,
построенного
из метрик $\mathsf{H}_{\ \nu
}^{\mu}=\mathsf{H}_{\ \nu}^{\mu}(\mathsf{g}^{\left(  1\right)  }%
,\ldots,\mathsf{g}^{\left(  N\right)  })$. В
этом случае
собственные
значения
матрицы $\mathsf{\hat{H}}$,
соответствующей
тензору $\mathsf{H}_{\ \nu}^{\mu}$,
являются
инвариантами
относительно
действия
общих
координатных
преобразований
$x^{\mu}\longmapsto\tilde{x}^{\mu}$,
поскольку $\dfrac
{\partial\tilde{x}^{\alpha}}{\partial x^{\mu}}\mathsf{H}_{\ \nu}^{\mu}%
\dfrac{\partial x^{\nu}}{\partial\tilde{x}^{\beta}}=\mathsf{\tilde{H}%
}_{\ \beta}^{\alpha}$.

Параметризуем
$\mathsf{\hat{H}}(\mathsf{g}^{\left(  1\right)  },\ldots,\mathsf{g}^{\left(
N\right)  })$, используя
следующее
замечание. В
большинстве
физически
интересных
моделей \cite{wald}
метрика
имеет
диагональную
форму, то
есть%
\begin{equation}
\mathsf{g}_{\mu\nu}^{\left(  i\right)  }=\mathrm{diag}\left(  \lambda
_{0}^{\left(  i\right)  },\lambda_{1}^{\left(  i\right)  },\ldots
,\lambda_{D-1}^{\left(  i\right)  }\right)  , \label{gd}%
\end{equation}
где $\lambda_{a}^{\left(  i\right)  }$
собственные
значения $i$%
-той метрики.
Следовательно,
структура
матрицы $\mathsf{\hat{H}}%
(\mathsf{g}^{\left(  1\right)  },\ldots,\mathsf{g}^{\left(  N\right)  })$
может быть
описана по
аналогии со
структурой
инвариантного
объема
взаимодействия
(построенного в разделе \ref{sec-inv}),
а именно,
построим $N$
матриц $\mathsf{H}_{\ \nu}^{\left(
i\right)  \mu}$ как
следующее
произведение
диагональных
метрик%
\begin{align}
\mathsf{H}_{\ \nu}^{\left(  i\right)  \mu}  &  =\mathsf{g}^{\left(  i\right)
\mu\alpha_{1}}\mathsf{g}_{\alpha_{1}\rho_{1}\ }^{\left(  1\right)  }%
\mathsf{g}^{\left(  i\right)  \rho_{1}\beta_{1}}\mathsf{g}_{\beta_{1}\rho_{2}%
}^{\left(  2\right)  }\cdot\ldots\cdot\mathsf{g}^{\left(  i\right)  \rho
_{j-1}\alpha_{j}}\mathsf{g}_{\alpha_{j}\rho_{j}\ }^{\left(  j\right)
}\mathsf{g}^{\left(  i\right)  \rho_{j}\beta_{j}}\mathsf{g}_{\beta_{j}%
\rho_{j+1}}^{\left(  j+1\right)  }\nonumber\\
&  \cdot\ldots\cdot\mathsf{g}^{\left(  i\right)  \rho_{N-2}\alpha_{N-1}%
}\mathsf{g}_{\alpha_{N-1}\rho_{N-1}\ }^{\left(  N-1\right)  }\mathsf{g}%
^{\left(  i\right)  \rho_{N-1}\beta_{N-1}}\mathsf{g}_{\beta_{N-1}\nu}^{\left(
N\right)  }.
\end{align}

Таким
образом
построенные
матрицы $\mathsf{\hat{H}}^{\left(
i\right)  }$
удовлетворяют
тождеству%
\begin{equation}
\mathsf{\hat{H}}^{\left(  1\right)  }\mathsf{\hat{H}}^{\left(  2\right)
}\ldots\mathsf{\hat{H}}^{\left(  N\right)  }=\mathsf{I}, \label{h11}%
\end{equation}
где $\mathsf{I}$ --- $D\times D$
единичная
матрица. Так
что имеется
$\left(  N-1\right)  $
независимых
матриц $\mathsf{\hat{H}}^{\left(
i\right)  }$. В случае
бигравитации
($N=2$) имеем две
матрицы
\begin{align}
\mathsf{H}_{\ \nu}^{\left(  1\right)  \mu}  &  =\mathsf{g}^{\left(  1\right)
\mu\beta_{1}}\mathsf{g}_{\beta_{1}\nu}^{\left(  2\right)  },\\
\mathsf{H}_{\ \nu}^{\left(  2\right)  \mu}  &  =\mathsf{g}^{\left(  2\right)
\mu\alpha_{1}}\mathsf{g}_{\alpha_{1}\nu}^{\left(  1\right)  },
\end{align}
которые
являются
взаимно
обратными $\mathsf{\hat
{H}}^{\left(  1\right)  }\mathsf{\hat{H}}^{\left(  2\right)  }=\mathsf{I}$
(см. (\ref{h11})),
поэтому
достаточно
рассматривать
только одну
из них (см.,
например, \cite{dam/kog}).
Исходя из
этого,
целесообразно
определить
следующие $N^{2}$
матриц $\mathsf{\hat{p}}^{\left(
i,j\right)  }$ как%
\begin{equation}
\mathsf{p}_{\ \ \ \ \text{\ }\nu}^{\left(  i,j\right)  \mu}=\mathsf{g}%
^{\left(  i\right)  \mu\rho}\mathsf{g}_{\rho\nu}^{\left(  j\right)  },
\label{h}%
\end{equation}
где $i,j=1,\ldots N$.
Очевидно,
что $p$-матрицы
$\mathsf{\hat{p}}^{\left(  i,j\right)  }$
удовлетворяют
соотношениям%
\begin{align}
\mathsf{\hat{p}}^{\left(  i,j\right)  }\mathsf{\hat{p}}^{\left(  j,k\right)
}  &  =\mathsf{\hat{p}}^{\left(  i,k\right)  },\label{hhh1}\\
\mathsf{\hat{p}}^{\left(  i,j\right)  }\mathsf{\hat{p}}^{\left(  j,i\right)
}  &  =\mathsf{\hat{p}}^{\left(  i,i\right)  }=\mathsf{I}. \label{hhh2}%
\end{align}

Произведение
(\ref{hhh1})
ассоциативно
и обратимо (\ref{hhh2}),
но
определено
не для всех
элементов,
поэтому
множество $p$%
-переменных
является
частичной
группой \cite{hermann}.
Заметим, что
существует
$N\left(  N-1\right)  \diagup2$
независимых
$p$-матриц,
которые
коммутируют
в случае
диагональных
метрик (\ref{gd}). В
бигравитации
$N=2$, имеем%
\begin{align}
\mathsf{\hat{H}}^{\left(  1\right)  }  &  =\mathsf{\hat{p}}^{\left(
1,2\right)  },\\
\mathsf{\hat{H}}^{\left(  2\right)  }  &  =\mathsf{\hat{p}}^{\left(
2,1\right)  }.
\end{align}

Построим
матрицы $\mathsf{\hat{H}}^{\left(
i\right)  }$ из 6
независимых
$p$-матриц \textsf{$p$}$^{\left(
i,j\right)  }$ для
случая
тернарной
гравитации ($N=3$)%
\begin{align}
\mathsf{\hat{H}}^{\left(  1\right)  }  &  =\mathsf{\hat{p}}^{\left(
1,3\right)  }\mathsf{\hat{p}}^{\left(  1,2\right)  },\\
\mathsf{\hat{H}}^{\left(  2\right)  }  &  =\mathsf{\hat{p}}^{\left(
2,1\right)  }\mathsf{\hat{p}}^{\left(  2,3\right)  },\\
\mathsf{\hat{H}}^{\left(  3\right)  }  &  =\mathsf{\hat{p}}^{\left(
3,2\right)  }\mathsf{\hat{p}}^{\left(  3,1\right)  },
\end{align}
которые
удовлетворяют
тождеству%
\begin{equation}
\mathsf{\hat{H}}^{\left(  1\right)  }\mathsf{\hat{H}}^{\left(  2\right)
}\mathsf{\hat{H}}^{\left(  3\right)  }=\mathsf{I}.
\end{equation}

Используя (\ref{gd}),
можно
представить
структуру
собственных
значений
матриц $\mathsf{H}_{\ \nu}^{\left(
i\right)  \mu}$ через
собственные
значения
метрик как%

\begin{equation}
\mathsf{\hat{H}}^{\left(  i\right)  }=\mathrm{diag}\left(  \dfrac{\left(
\lambda_{0}^{\left(  i\right)  }\right)  ^{N}}{R_{0}},\dfrac{\left(
\lambda_{1}^{\left(  i\right)  }\right)  ^{N}}{R_{1}},\ldots,\dfrac{\left(
\lambda_{D-1}^{\left(  i\right)  }\right)  ^{N}}{R_{D-1}}\right)  ,
\label{hhh}%
\end{equation}
где $R_{a}=\Pi_{i=1}^{N}\lambda_{a}^{\left(  i\right)  }$.
Тогда из (\ref{hhh}),
следует что%
\begin{equation}
\det\mathsf{\hat{H}}^{\left(  i\right)  }=\dfrac{\left(  \det\mathsf{g}%
^{\left(  i\right)  }\right)  ^{N}}{\Pi_{j=1}^{N}\det\mathsf{g}^{\left(
j\right)  }}, \label{deth}%
\end{equation}
и, очевидно,
что $\Pi_{j=1}^{N}\det\mathsf{\hat{H}}^{\left(  j\right)
}=1$ (см. (\ref{h11})).

Отметим, что
для метрики
$\mathsf{g}_{\mu\nu}^{\left(  i\right)  }$ с
сигнатурой
$\left(  +,\overset{D-1}{\overbrace{-,\ldots-}}\right)  $
знаки
собственных
чисел
определены
так
 (см., например, \cite{wald}) $\lambda_{0}^{\left(
i\right)  }>0$, $\lambda_{1}^{\left(  i\right)  }<0$, $\ldots$, $\lambda
_{D-1}^{\left(  i\right)  }<0$.
Учитывая (\ref{hhh}) и
(\ref{h11}), получаем,
что все
собственные
значения
матриц $\mathsf{\hat{H}}^{\left(
i\right)  }$ являются
положительными
и
ненулевыми.
Это
позволяет
определить
новые $\mu$%
-переменные%
\begin{equation}
\mu_{a}^{\left(  i\right)  }=\ln\dfrac{\left(  \lambda_{a}^{\left(  i\right)
}\right)  ^{N}}{R_{a}}, \label{ml}%
\end{equation}
которые
удовлетворяют
$D$ тождествам%
\begin{equation}
\sum_{i=1}^{N}\mu_{a}^{\left(  i\right)  }=0,\ \ \ a=0,\ldots,D-1. \label{m0}%
\end{equation}

Учитывая (\ref{m0}),
число
независимых
$\mu$-переменных
есть $D\left(  N-1\right)  $.
Таким
образом,
скалярный
потенциал
взаимодействия
может быть
выбран как
гладкая
функция $\mu$%
-переменных,
т.е. \begin{equation}V(\mathsf{g}^{\left(  1\right)  },\mathsf{g}^{\left(
2\right)  },\ldots,\mathsf{g}^{\left(  N\right)  })=\tilde{v}\left(  \mu
_{a}^{\left(  i\right)  }\right)  .\end{equation}
Следуя \cite{dam/kog}
(где
рассматривался
частный
случай $N=2$, $D=4$),
выбираем более удобный 
базис в
виде
симметричных
полиномов%
\begin{equation}
\sigma_{k}^{\left(  i\right)  }=\sum_{a=0}^{D-1}\left(  \mu_{a}^{\left(
i\right)  }\right)  ^{k},\;\; k=1,\ldots,D, \label{sk}%
\end{equation}
связанных между собой $D$ соотношениями, 
следующими из
(\ref{m0}). Таким
образом,
скалярный
потенциал
взаимодействия
для
мультигравитации
можно записать в симметричном виде%
\begin{equation}
V(\mathsf{g}^{\left(  1\right)  },\mathsf{g}^{\left(
2\right)  },\ldots,\mathsf{g}^{\left(  N\right)  })=v\left(  \sigma_{1}^{\left(  i\right)
},\sigma_{2}^{\left(  i\right)  },\ldots\sigma_{D}^{\left(  i\right)
}\right)  ,\ \ \ i=1,\ldots,N, \label{vg}%
\end{equation}
где $v$ ---
скалярная
функция от $D\left( N-1 \right)$ независимых полиномов
$\sigma_{k}^{\left(  i\right)  }$.

Мы естественно предполагаем,
что в случае
плоских
пространств
взаимодействие
отсутствует.
Тогда имеем 
\textquotedblleft%
граничное\textquotedblright%
\  условие%
\begin{equation}
v\left(  0,0,\ldots0\right)  =0. \label{v0}%
\end{equation}

Выразим
скалярный
потенциал
взаимодействия
(\ref{vg}) через
комбинацию
инвариантов
матриц $\mathsf{\hat{H}}^{\left(
i\right)  }$ в явном
виде. Из (\ref{deth}), (\ref{ml}) и
(\ref{sk}), получаем%
\begin{equation}
\sigma_{k}^{\left(  i\right)  }=\operatorname*{tr}\left(  \ln\mathsf{\hat{H}%
}^{\left(  i\right)  }\right)  ^{k}. \label{st}%
\end{equation}

Параметризуем
метрику как%
\begin{equation}
\mathsf{g}_{\mu\nu}^{\left(  i\right)  }=\eta_{\mu\nu}+\mathsf{h}_{\mu\nu
}^{\left(  i\right)  }, \label{gh}%
\end{equation}
где $\mathsf{h}_{\mu\nu}^{\left(  i\right)  }$
некоторые
возмущения
над плоским
фоном.
Ограничиваясь
квадратичными
слагаемыми
по
возмущениям
$\mathsf{h}_{\mu\nu}^{\left(  i\right)  }$,
которые
соответствуют
массивному
случаю и
отсутствию
самодействия,
получаем для
$\sigma_{1}^{\left(  i\right)  }$ и $\sigma_{2}^{\left(  i\right)  }$
следующие
выражения%
\begin{align}
\sigma_{1}^{\left(  i\right)  }  &  =\sum_{\substack{j=1\\j\neq i}}^{N}\left[
\left(  h^{\left(  i\right)  }-h^{\left(  j\right)  }\right)  -\left(  \left(
h_{\mu\nu}^{\left(  i\right)  }\right)  ^{2}-\left(  h_{\mu\nu}^{\left(
j\right)  }\right)  ^{2}\right)  \right]  ,\label{s1}\\
\sigma_{2}^{\left(  i\right)  }  &  =\left(  N-1\right)  ^{2}\left(  h_{\mu
\nu}^{\left(  i\right)  }\right)  ^{2}+\sum_{\substack{j=1\\j\neq i}%
}^{N}\left(  h_{\mu\nu}^{\left(  j\right)  }\right)  ^{2}\nonumber\\
&  +2\sum_{\substack{k,j=1\\j\neq k,k\neq i,j\neq i}}^{N}\mathsf{h}%
_{\ \ \ \ \nu}^{\left(  j\right)  \mu}\mathsf{h}_{\ \ \ \ \mu}^{\left(
k\right)  \nu}-2\left(  N-1\right)  \mathsf{h}_{\ \ \ \ \nu}^{\left(
i\right)  \mu}\sum_{\substack{j=1\\j\neq i}}^{N}\mathsf{h}_{\ \ \ \ \mu
}^{\left(  j\right)  \nu}. \label{s2}%
\end{align}
где $h^{\left(  i\right)  }:=\mathsf{h}_{\mu\nu}^{\left(
i\right)  }\eta^{\mu\nu}$ и $\left(  h_{\mu\nu}^{\left(  i\right)
}\right)  ^{2}:=\mathsf{h}_{\mu\nu}^{\left(  i\right)  }\mathsf{h}^{\left(
i\right)  \mu\nu}$. Заметим,
что $\sigma_{k}^{\left(  i\right)  }\sim\mathcal{O}\left(
\left(  h^{\left(  i\right)  }\right)  ^{k}\right)  $,
следовательно,
ограничиваясь
квадратичными
слагаемыми,
нет
необходимости
рассматривать
выражения со
степенями $k\geq3$.

Следовательно,
скалярный
потенциал
взаимодействия
в
мультигравитации
в
квадратичном
приближении
можно представить в виде%
\begin{equation}
V(\mathsf{g}^{\left(  i\right)  })=\sum_{i=1}^{N}\left[  a_{i}\sigma
_{1}^{\left(  i\right)  }+b_{i}\left(  \sigma_{1}^{\left(  i\right)  }\right)
^{2}+c_{i}\sigma_{2}^{\left(  i\right)  }\right]  , \label{vgi}%
\end{equation}
где $a_{i},b_{i},c_{i}$
произвольные
действительнве
константы.
Из (\ref{s1}) следует
что%
\begin{equation}
\sum_{i=1}^{N}\sigma_{1}^{\left(  i\right)  }=0,
\end{equation}
как и должно
быть из (\ref{m0}).

\section{МОДЕЛЬ
ПАУЛИ-ФИРЦА
В
БИГРАВИТАЦИИ}%

В качестве
примера
рассмотрим
бигравитацию
($N=2$) и получим
из общих
принципов
модель
Паули-Фирца.
Так, вместо
(\ref{s1})--(\ref{s2}) имеем (с
точностью до
квадратичных
слагаемых по
возмущениям
$\mathsf{h}_{\mu\nu}^{\left(  1,2\right)  }$)%
\begin{align}
\sigma_{1}^{\left(  1\right)  }  &  =-\sigma_{1}^{\left(  2\right)
}=h^{\left(  1\right)  }-h^{\left(  2\right)  }-\left(  \left(  h_{\mu\nu
}^{\left(  1\right)  }\right)  ^{2}-\left(  h_{\mu\nu}^{\left(  2\right)
}\right)  ^{2}\right)  \equiv\sigma_{1},\\
\sigma_{2}^{\left(  1\right)  }  &  =\sigma_{2}^{\left(  2\right)  }=\left(
h_{\mu\nu}^{\left(  1\right)  }\right)  ^{2}+\left(  h_{\mu\nu}^{\left(
2\right)  }\right)  ^{2}-2\mathsf{h}_{\ \ \ \ \nu}^{\left(  1\right)  \mu
}\mathsf{h}_{\ \ \ \ \mu}^{\left(  2\right)  \nu}\equiv\sigma_{2}.
\end{align}

Для
скалярного
потенциала
взаимодействия
сумма (\ref{vgi})
принимает
вид (с учетом
(\ref{v0}))%
\begin{equation}
V(\mathsf{g}^{\left(  1\right)  },\mathsf{g}^{\left(  2\right)  })=a\sigma
_{1}+b\sigma_{1}^{2}+c\sigma_{2}, \label{v12}%
\end{equation}
где $a,b,c$ -
произвольные
действительные
константы
размерности
$(mass)^{4}$. Тогда
полное
действие для
бигравитации
запишется
как%
\begin{equation}
S_{2}=-M_{1}^{2}\int d^{4}xR_{1}\sqrt{g_{1}}-M_{2}^{2}\int d^{4}xR_{2}%
\sqrt{g_{2}}+\int d\Omega_{int}^{\left(  2\right)  }V(\mathsf{g}^{\left(
1\right)  },\mathsf{g}^{\left(  2\right)  }), \label{a2}%
\end{equation}
где $M_{1,2}$
константы
размерности
$(mass)^{1}$, и $d\Omega_{int}^{\left(  2\right)  }$ -
инвариантный
объем
взаимодействия
для
бигравитации
(\ref{w}), который
имеет вид%
\begin{equation}
d\Omega_{int}^{\left(  2\right)  }=d^{4}x\cdot\sqrt[4]{g_{1}g_{2}}\cdot
\frac{1}{\alpha+\beta+\gamma}\left[  \frac{\alpha}{2}\left(  \sqrt
{\dfrac{g_{1}}{g_{2}}}+\sqrt{\dfrac{g_{2}}{g_{1}}}\right)  +\beta
+2\gamma\left(  \sqrt{\dfrac{g_{1}}{g_{2}}}+\sqrt{\dfrac{g_{2}}{g_{1}}%
}\right)  ^{-1}\right]  , \label{dom2}%
\end{equation}
где $\alpha,\beta,\gamma$ -
безразмерные
параметры и
$\alpha+\beta+\gamma\neq0$. Заметим,
что
параметризация
(\ref{gh}) выражения
(\ref{dom2}) приводит к
виду%
\begin{equation}
d\Omega_{int}^{\left(  2\right)  }=d^{4}x\cdot\sqrt[4]{g_{1}g_{2}}+\ldots,
\end{equation}
где $\ldots$
означают
слагаемые
квадратичные
по
возмущениям
$\mathsf{h}_{\mu\nu}^{\left(  1,2\right)  }$. Эти
слагаемые не
вносят
вклада в (\ref{a2}),
потому что
мы
ограничиваемся
вторым
порядком, а
скалярный
потенциал
взаимодействия
(\ref{v12}) не
содержит
слагаемых
без $\mathsf{h}_{\mu\nu}^{\left(  1,2\right)  }$.
Используя
разложение
(\ref{gh}) и применяя
его к
действию (\ref{a2}),
получаем%
\begin{equation}
S_{2}=\int d^{4}x\left(  L_{kin}+L_{int}\right)  , \label{s2x}%
\end{equation}
где%
\begin{align}
L_{kin}  &  =\dfrac{1}{4}M_{1}^{2}\left[  \partial^{\rho}\mathsf{h}_{\mu\nu
}^{\left(  1\right)  }\partial_{\rho}\mathsf{h}^{\left(  1\right)  \mu\nu
}-\partial^{\mu}h^{\left(  1\right)  }\partial_{\mu}h^{\left(  1\right)
}+2\partial_{\mu}\mathsf{h}^{\left(  1\right)  \mu\nu}\partial_{\nu}h^{\left(
1\right)  }-2\partial_{\mu}\mathsf{h}^{\left(  1\right)  \mu\nu}\partial
_{\rho}\mathsf{h}_{\nu}^{\left(  1\right)  \rho}\right] \nonumber\\
&  +\dfrac{1}{4}M_{2}^{2}\left[  \partial^{\rho}\mathsf{h}_{\mu\nu}^{\left(
2\right)  }\partial_{\rho}\mathsf{h}^{\left(  2\right)  \mu\nu}-\partial^{\mu
}h^{\left(  2\right)  }\partial_{\mu}h^{\left(  2\right)  }+2\partial_{\mu
}\mathsf{h}^{\left(  2\right)  \mu\nu}\partial_{\nu}h^{\left(  2\right)
}-2\partial_{\mu}\mathsf{h}^{\left(  2\right)  \mu\nu}\partial_{\rho
}\mathsf{h}_{\nu}^{\left(  2\right)  \rho}\right]  ,
\end{align}

\begin{align}
L_{int}  &  =a(h^{\left(  1\right)  }-h^{\left(  2\right)  })^{2}+b\left(
\mathsf{h}_{\mu\nu}^{\left(  1\right)  }-\mathsf{h}_{\mu\nu}^{\left(
2\right)  }\right)  (\mathsf{h}^{\left(  1\right)  \mu\nu}-\mathsf{h}^{\left(
2\right)  \mu\nu})+c\left(  \mathsf{h}_{\mu\nu}^{\left(  2\right)  }%
\mathsf{h}^{\left(  2\right)  \mu\nu}-\mathsf{h}_{\mu\nu}^{\left(  1\right)
}\mathsf{h}^{\left(  1\right)  \mu\nu}\right) \nonumber\\
&  +\dfrac{c}{4}\left(  \left(  h^{\left(  1\right)  }\right)  ^{2}-\left(
h^{\left(  2\right)  }\right)  ^{2}\right)  .
\end{align}

Далее,
применим
(3+1)-разложение
\cite{rub/tin} для
полного
действия (\ref{s2x}).
Отделим
пространственные
и временные
компоненты в
$L_{int}$, тогда%

\begin{align}
L_{int}  &  =a\left(  h_{00}^{\left(  1\right)  }-h_{00}^{\left(  2\right)
}-h_{ii}^{\left(  1\right)  }+h_{ii}^{\left(  2\right)  }\right)
^{2}+\nonumber\\
&  +b\left(  h_{00}^{\left(  1\right)  }-h_{00}^{\left(  2\right)  }\right)
\left(  h_{00}^{\left(  1\right)  }-h_{00}^{\left(  2\right)  }\right)
-2b\left(  h_{0i}^{\left(  1\right)  }-h_{0i}^{\left(  2\right)  }\right)
\left(  h_{0i}^{\left(  1\right)  }-h_{0i}^{\left(  2\right)  }\right)
+b\left(  h_{ij}^{\left(  1\right)  }-h_{ij}^{\left(  2\right)  }\right)
\left(  h_{ij}^{\left(  1\right)  }-h_{ij}^{\left(  2\right)  }\right)
+\nonumber\\
&  +c\left(  h_{00}^{\left(  2\right)  }h_{00}^{\left(  2\right)  }%
-2h_{0i}^{\left(  2\right)  }h_{0i}^{\left(  2\right)  }+h_{ij}^{\left(
2\right)  }h_{ij}^{\left(  2\right)  }-h_{00}^{\left(  1\right)  }%
h_{00}^{\left(  1\right)  }+2h_{0i}^{\left(  1\right)  }h_{0i}^{\left(
1\right)  }-h_{ij}^{\left(  1\right)  }h_{ij}^{\left(  1\right)  }\right)
+\nonumber\\
&  +\dfrac{c}{4}\left(  \left(  h_{00}^{\left(  1\right)  }-h_{ii}^{\left(
1\right)  }\right)  ^{2}-\left(  h_{00}^{\left(  2\right)  }-h_{ii}^{\left(
2\right)  }\right)  ^{2}\right)  \label{s22}%
\end{align}

Мы
ограничемся
рассмотрением
только
скалярного
сектора, так
как это
является
вполне
достаточным
для
уничтожения
духовых мод
в спектре
(для
стандартной
гравитации,
см. \cite{rub/tin}).
Параметризуем
(3+1) разложение
в виде%
\begin{align}
h_{00}^{(1,2)}  &  =2\varphi_{(1,2)},\\
h_{0i}^{(1,2)}  &  =\partial_{i}B_{(1,2)},\\
h_{ij}^{(1,2)}  &  =-2(\psi_{(1,2)}\delta_{ij}-\partial_{i}\partial
_{j}E_{(1,2)}),
\end{align}
где $\varphi_{(1,2)},\psi_{(1,2)},B_{(1,2)},E_{(1,2)}$ ---
скалярные
поля для
возмущеной
метрики $\mathsf{h}_{\mu\nu
}^{\left(  1\right)  }$ и $\mathsf{h}_{\mu\nu}^{\left(  2\right)  }$
соответственно.
Из (\ref{s2x})
получаем для
кинетического
слагаемого%
\begin{align}
L_{kin}  &  =M_{1}^{2}\left[  -2\psi_{1}\partial_{k}\partial_{k}\psi_{1}%
-6\dot{\psi}_{1}^{2}-4\varphi_{1}\partial_{k}\partial_{k}\psi_{1}-4\dot{\psi
}_{1}\partial_{k}\partial_{k}B_{1}+4\dot{\psi}_{1}\partial_{k}\partial_{k}%
\dot{E}_{1}\right]  +\nonumber\\
&  +M_{2}^{2}\left[  -2\psi_{2}\partial_{k}\partial_{k}\psi_{2}-6\dot{\psi
}_{2}^{2}-4\varphi_{2}\partial_{k}\partial_{k}\psi_{2}-4\dot{\psi}_{2}%
\partial_{k}\partial_{k}B_{2}+4\dot{\psi}_{2}\partial_{k}\partial_{k}\dot
{E}_{2}\right]  , \label{lkin}%
\end{align}
и
взаимодействия%
\begin{align}
L_{int}  &  =a\left(  2(\varphi_{1}-\varphi_{2})+6(\psi_{1}-\psi_{2}%
)-2\Delta(E_{1}-E_{2})\right)  ^{2}+\nonumber\\
&  +b\left(  4(\varphi_{1}-\varphi_{2})^{2}+2(B_{1}-B_{2})(\Delta B_{1}-\Delta
B_{2})+12(\psi_{1}-\psi_{2})^{2}+4(\Delta E_{1}-\Delta E_{2})^{2}\right.
-\nonumber\\
&  \left.  -8(\psi_{1}-\psi_{2})(\Delta E_{1}-\Delta E_{2})\right)
+\nonumber\\
&  +c\left(  4\left(  \varphi_{2}^{2}-\varphi_{1}^{2}\right)  +12\left(
\psi_{2}^{2}-\psi_{1}^{2}\right)  +B_{2}\Delta B_{2}-B_{1}\Delta
B_{1}+4\left(  \left(  \Delta E_{2}\right)  ^{2}-\left(  \Delta E_{1}\right)
^{2}\right)  \right.  +\nonumber\\
&  \left.  +8\left(  \psi_{1}\Delta E_{1}-\psi_{2}\Delta E_{2}\right)
\right)  +c\left(  \left(  \varphi_{1}+3\psi_{1}-\Delta E_{1}\right)
^{2}-\left(  \varphi_{2}+3\psi_{2}-\Delta E_{2}\right)  ^{2}\right)  .
\label{lint}%
\end{align}

Далее,
рассматрим
часть
полного
лагранжиана,
содержащую
скалярные
поля $\varphi_{(1,2)}$,%
\begin{align}
L(\varphi)  &  =-4M_{1}^{2}\varphi_{1}\Delta\psi_{1}-4M_{2}^{2}\varphi
_{2}\Delta\psi_{2}+\varphi_{1}^{2}\left(  4a+4b-3c\right)  +\varphi_{2}%
^{2}\left(  4a+4b+3c\right)  +\nonumber\\
&  +\varphi_{1}\left(  24a\left(  \psi_{1}-\psi_{2}\right)  -8a\left(  \Delta
E_{1}-\Delta E_{2}\right)  +6c\psi_{1}-2c\Delta E_{1}\right)  +\nonumber\\
&  +\varphi_{2}\left(  -24a\left(  \psi_{1}-\psi_{2}\right)  +8a\left(  \Delta
E_{1}-\Delta E_{2}\right)  -6c\psi_{2}+2c\Delta E_{2}\right)  -8\varphi
_{1}\varphi_{2}\left(  a+b\right)  .
\end{align}

Очевидно,
что при
\begin{align}
4a+4b-3c  &  =0\label{a4}\\
4a+4b+3c  &  =0\\
a+b  &  =0 \label{ab}%
\end{align}
лагранжиан
не содержит
квадратичных
слагаемых по
полям $\varphi_{(1,2)}$, то
есть
скалярные
поля
являются
нединамическими
(подробнее
см. \cite{rub/tin}).
Система (\ref{a4})--(\ref{ab}) эквивалентна%

\begin{equation}
a+b=0,\ \ \ \ \ c=0 \label{abc}%
\end{equation}

Отметим, что
только при
таких
соотношениях
на параметры
лагранжиан
можно
представить
через
разности
соответствующих
полей. Введем%

\begin{align}
\varphi &  \equiv\varphi_{1}-\varphi_{2}\label{fi}\\
B  &  \equiv B_{1}-B_{2}\label{b1}\\
\psi &  \equiv\psi_{1}-\psi_{2}\label{p12}\\
E  &  \equiv E_{1}-E_{2} \label{EE1-E2}%
\end{align}
тогда
лагранжиан
взаимодействия
(\ref{lint}) принимает
вид%
\begin{equation}
L_{int}^{(2)}=4a\left[  6\psi^{2}+6\varphi\psi-2\varphi\Delta E-4\psi\Delta
E-\dfrac{1}{2}B\Delta B\right]
\end{equation}
Данное
выражение
совпадает с
массовым
лагранжианом
Паули-Фирца
в 3+1
разложении
стандартной
гравитации
\cite{rub/tin}. Для того,
чтобы
доказать
эквивалентность
бигравитации
(\ref{a2}) и теории
Паули-Фирца,
необходимо
включить в
рассмотрение
также и
кинетическую
часть.
Отметим, что
кинетическое
слагаемое (\ref{lkin})
можно
представить
через поля
(\ref{fi})--(\ref{EE1-E2}) только с
использованием
уравнений
движения.
Для этого
выпишем
полный
лагранжиан
(\ref{lkin}) и (\ref{lint}) с
учетом (\ref{fi}) и (\ref{b1}),
имеем%
\begin{align}
L_{kin}^{(2)}+L_{int}^{(2)}  &  =M_{1}^{2}\left[  -2\psi_{1}\partial
_{k}\partial_{k}\psi_{1}-6\dot{\psi}_{1}^{2}-4\varphi_{1}\partial_{k}%
\partial_{k}\psi_{1}-4\dot{\psi}_{1}\partial_{k}\partial_{k}B_{1}+4\dot{\psi
}_{1}\partial_{k}\partial_{k}\dot{E}_{1}\right]  +\nonumber\\
&  +M_{2}^{2}\left[  -2\psi_{2}\partial_{k}\partial_{k}\psi_{2}-6\dot{\psi
}_{2}^{2}-4\varphi_{2}\partial_{k}\partial_{k}\psi_{2}-4\dot{\psi}_{2}%
\partial_{k}\partial_{k}B_{2}+4\dot{\psi}_{2}\partial_{k}\partial_{k}\dot
{E}_{2}\right]  +\nonumber\\
&  +24a\left(  \psi_{1}-\psi_{2}\right)  ^{2}+4a\left[  6\left(  \varphi
_{1}-\varphi_{2}\right)  \left(  \psi_{1}-\psi_{2}\right)  -2\left(
\varphi_{1}-\varphi_{2}\right)  \Delta\left(  E_{1}-E_{2}\right)  \right]
-\nonumber\\
&  -16a\left(  \psi_{1}-\psi_{2}\right)  \Delta\left(  E_{1}-E_{2}\right)
-2a\left(  B_{1}-B_{2}\right)  \Delta\left(  B_{1}-B_{2}\right)
\label{L_EH_L_int}%
\end{align}

Система
уравнений
Эйлера-Лагранжа
по полям $B_{1,2}$
принимает
вид%
\begin{equation}
\left\{
\begin{array}
[c]{c}%
4M_{1}^{2}\Delta\dot{\psi}_{1}+4a\left(  \Delta B_{1}-\Delta B_{2}\right)
=0\\
4M_{2}^{2}\Delta\dot{\psi}_{2}+4a\left(  \Delta B_{2}-\Delta B_{1}\right)  =0
\end{array}
\right.  , \label{m12}%
\end{equation}
где мы
представили
нужную нам
часть
лагранжиана
как%
\begin{equation}
L(B)=4M_{1}^{2}\partial_{k}\dot{\psi}_{1}\partial_{k}B_{1}+4M_{2}^{2}%
\partial_{k}\dot{\psi}_{2}\partial_{k}B_{2}+2a\left(  \partial_{k}%
B_{1}-\partial_{k}B_{2}\right)  \left(  \partial_{k}B_{1}-\partial_{k}%
B_{2}\right)  . \label{L_B}%
\end{equation}

Учитывая (\ref{b1})
формула (\ref{m12})
переходит в%
\begin{equation}
\left\{
\begin{array}
[c]{c}%
M_{1}^{2}\Delta\dot{\psi}_{1}=-a\Delta B\\
M_{2}^{2}\Delta\dot{\psi}_{2}=a\Delta B
\end{array}
\right.  , \label{psiB}%
\end{equation}
откуда
следует%
\begin{equation}
M_{1}^{2}\psi_{1}=-M_{2}^{2}\psi_{2}. \label{psi1_psi2}%
\end{equation}
Для поля $\psi$ см.
(\ref{p12}) получаем%
\begin{equation}
\psi=\psi_{1}-\psi_{2}=\psi_{1}+\dfrac{M_{1}^{2}}{M_{2}^{2}}\psi_{1}%
=\dfrac{M_{1}^{2}+M_{2}^{2}}{M_{2}^{2}}\psi_{1}=-\dfrac{M_{2}^{2}}{M_{1}^{2}%
}\psi_{2}-\psi_{2}=-\dfrac{M_{1}^{2}+M_{2}^{2}}{M_{1}^{2}}\psi_{2}.
\end{equation}
Учитывая (\ref{psiB}),
имеем%
\begin{equation}
B=\dfrac{M_{1}^{2}}{-a}\dot{\psi}_{1}=\dfrac{M_{1}^{2}M_{2}^{2}}{-a\left(
M_{1}^{2}+M_{2}^{2}\right)  }\dot{\psi}%
\end{equation}
Тогда, часть
лагранжиана
$L(B),$ (\ref{L_B})
принимает вид%

\begin{equation}
L(B)=2\dfrac{M_{1}^{4}M_{2}^{4}}{a\left(  M_{1}^{2}+M_{2}^{2}\right)  ^{2}%
}\dot{\psi}\Delta\dot{\psi}.
\end{equation}
Варьирование
(\ref{L_EH_L_int}) по полям
$\varphi_{(1,2)}$ приводит
к системе%
\begin{equation}
\left\{
\begin{array}
[c]{c}%
-M_{1}^{2}\Delta\psi_{1}+6a\left(  \psi_{1}-\psi_{2}\right)  -2a\left(  \Delta
E_{1}-\Delta E_{2}\right)  =0\\
-M_{2}^{2}\Delta\psi_{2}-6a\left(  \psi_{1}-\psi_{2}\right)  +2a\left(  \Delta
E_{1}-\Delta E_{2}\right)  =0
\end{array}
\right.  ,
\end{equation}
которая, с
учетом (\ref{EE1-E2}) и
(\ref{psi1_psi2}),
эквивалентна
выражению%

\begin{equation}
\Delta E=-\dfrac{M_{1}^{2}M_{2}^{2}}{2a\left(  M_{1}^{2}+M_{2}^{2}\right)
}\Delta\psi+3\psi.
\end{equation}

Тогда, часть
лагранжиана,
содержащую
поля $E_{1,2},$ можно
переписать в
виде%
\begin{align}
L(E)  &  =4M_{1}^{2}\dot{\psi}_{1}\Delta\dot{E}_{1}+4M_{2}^{2}\dot{\psi}%
_{2}\Delta\dot{E}_{2}-8a\varphi\Delta E-16a\psi\Delta E=\\
&  =4\dfrac{M_{1}^{2}M_{2}^{2}}{M_{1}^{2}+M_{2}^{2}}\dot{\psi}\left(
-\dfrac{M_{1}^{2}M_{2}^{2}}{2a\left(  M_{1}^{2}+M_{2}^{2}\right)  }\Delta
\dot{\psi}+3\dot{\psi}\right)  -8a\left(  \varphi+2\psi\right)  \left(
-\dfrac{M_{1}^{2}M_{2}^{2}}{2a\left(  M_{1}^{2}+M_{2}^{2}\right)  }\Delta
\psi+3\psi\right)  .
\end{align}

Оставшиеся
слагаемые в
кинетическом
выражении
полного
лагранжиана
(\ref{L_EH_L_int}) также
выразим
через поле $\psi$%
\begin{align}
L_{k}(\psi)  &  =-2M_{1}^{2}\psi_{1}\Delta\psi_{1}-2M_{2}^{2}\psi_{2}%
\Delta\psi_{2}-6M_{1}^{2}\dot{\psi}_{1}^{2}-6M_{2}^{2}\dot{\psi}_{2}%
^{2}-4M_{1}^{2}\varphi_{1}\Delta\psi_{1}-4M_{2}^{2}\varphi_{2}\Delta\psi
_{2}=\\
&  =-2\dfrac{M_{1}^{2}M_{2}^{2}}{M_{1}^{2}+M_{2}^{2}}\left(  \psi\Delta
\psi+3\dot{\psi}^{2}+2\varphi\Delta\psi\right)  .
\end{align}

Полный
лагранжиан
(\ref{L_EH_L_int}) есть%
\begin{align}
L_{EH}^{(2)}+L_{int}^{(2)}  &  =L_{k}(\psi)+L(B)+L(E)+24a\psi^{2}%
+24a\varphi\psi=\nonumber\\
&  =6\dfrac{M_{1}^{2}M_{2}^{2}}{M_{1}^{2}+M_{2}^{2}}\left(  \dot{\psi}%
^{2}+\psi\Delta\psi\right)  -24a\psi^{2}.
\end{align}

Представим
постоянную $a$
через новую
постоянную
$m_{g}^{2}$
\begin{equation}
a=\dfrac{1}{4}\dfrac{M_{1}^{2}M_{2}^{2}}{M_{1}^{2}+M_{2}^{2}}m_{g}^{2},
\end{equation}
тогда
скалярный
сектор
бигравитации
принимает
вид%
\begin{equation}
L=6\dfrac{M_{1}^{2}M_{2}^{2}}{M_{1}^{2}+M_{2}^{2}}\left(  \dot{\psi}^{2}%
+\psi\Delta\psi-m_{g}^{2}\psi^{2}\right)  =6\dfrac{M_{1}^{2}M_{2}^{2}}%
{M_{1}^{2}+M_{2}^{2}}\left(  \partial_{\mu}\psi\partial^{\mu}\psi-m_{g}%
^{2}\psi^{2}\right)  ,
\end{equation}
где $m_{g}$ - масса
гравитона.
Тогда
действие (\ref{a2}) с
учетом
условий (\ref{abc})
запишется,
как%
\begin{equation}
S_{g}=-M_{1}^{2}\int R_{1}\sqrt{-g_{1}}d^{4}x-M_{2}^{2}\int R_{2}\sqrt{-g_{2}%
}d^{4}x-\dfrac{1}{4}\dfrac{M_{1}^{2}M_{2}^{2}}{M_{1}^{2}+M_{2}^{2}}\int\left(
g_{1}g_{2}\right)  ^{1/4}d^{4}x(\sigma_{2}-\sigma_{1}^{2}).
\end{equation}

Из этого
следует, что
только
полное
действие
бигравитации
приводит к
теории
Паули-Фирца.
Отметим, что
в работе \cite{dam/kog}
слагаемое
взаимодействия
было
предложено
на основе
полу-эвристических
рассуждений,
в то время,
как мы
показали это 
в рамках квадратичного приближения, используя 3+1 разложение.

\section{ВЫВОДЫ}

Таким
образом, в
данной
работе мы
построили
инвариантный
объем
взаимодействия
мультигравитации
в общем виде.
Частный
случай
объема как
сумма трех
различных
средних (в
работе \cite{dam/kog}
рассматривалось
только
геометрическое
среднее) был
использован
при анализе
модели
бигравитации.
В рамках
формализма 3+1
разложения
нами строго
доказана
(в квадратичном приближении)
эквивалентность
полного
лагранжиана
бигравитации
(с учетом
кинетических
слагаемых
типа
эйнштейновских)
и массивной
теории Паули-Фирца.

\appendix

\section{Приложение. РАЗЛОЖНЕНИЕ
$\sqrt{g}$ ПО МАЛЫМ
ВОЗМУЩЕНИЯМ}

При
разложении
$\sqrt{g}$ по малым
возмущениям
$h_{\mu\nu}$
стандартным
образом
используем
выражение $\ln
(\det\mathsf{g}_{\mu\nu})=tr(\ln\mathsf{g}_{\mu\nu})$, из
которого
получаем
\begin{equation}
\sqrt{g}=exp\left(  \frac{1}{2}tr(\ln\mathsf{g}_{\mu\nu})\right)  .
\end{equation}
Над плоской
фоновой
метрикой $\mathsf{g}_{\mu\nu
}=\eta_{\mu\nu}+\mathsf{h}_{\mu\nu}$ имеем%
\begin{equation}
\sqrt{g}=1+\frac{1}{2}h-\frac{1}{4}\mathsf{h}_{\mu\alpha}\mathsf{h}^{\mu
\alpha}+\frac{1}{8}h^{2}, \label{sqrt}%
\end{equation}
с точностью
$O(h^{2})$, где $h=\mathsf{h}_{\mu\nu}\eta^{\mu\nu}$.

Приведем
здесь
методику
вычисления
разложения
$\sqrt{g}$ пригодную
для любой $\,^{(0)}%
\mathsf{g}_{\mu\nu}$ фоновой
метрики.
Общие
формулы для
разложения
$\sqrt{g}$ с
точностью до
первого
порядка были
приведены в
\cite{dup/kotv2007}, а с
точностью до
второго
порядка в \cite{kot/kru}.
Имеем%
\begin{equation}
\mathsf{g}_{\mu\nu}=\,^{(0)}\mathsf{g}_{\mu\nu}+h_{\mu\nu}.
\end{equation}
Тогда (в
случае $D=4$)
получаем%
\begin{equation}
\det\left(  ^{(0)}\mathsf{g}_{\mu\nu}+\mathsf{h}_{\mu\nu}\right)
=\varepsilon^{\alpha\beta\rho\sigma}\left(  ^{(0)}\mathsf{g}_{0\alpha
}+\mathsf{h}_{0\alpha}\right)  \left(  ^{(0)}\mathsf{g}_{1\beta}%
+\mathsf{h}_{1\beta}\right)  \left(  ^{(0)}\mathsf{g}_{2\rho}+\mathsf{h}%
_{2\rho}\right)  \left(  ^{(0)}\mathsf{g}_{3\sigma}+\mathsf{h}_{3\sigma
}\right)  ,
\end{equation}
где $\varepsilon^{0123}=+1$. С
точностью $O(h^{2})$
имеем%
\begin{equation}
\det\left(  ^{(0)}\mathsf{g}_{\mu\nu}+\mathsf{h}_{\mu\nu}\right)  =\det\left(
^{(0)}\mathsf{g}_{\mu\nu}\right)  +\mathsf{h}_{\mu\nu}\mathsf{K}^{\mu\nu
}\left(  ^{\left(  0\right)  }\mathsf{g}\right)  +\mathsf{h}_{\mu\nu
}\mathsf{h}_{\alpha\beta}\mathsf{F}^{\mu\nu\alpha\beta}\left(  ^{\left(
0\right)  }\mathsf{g}\right)  ,
\end{equation}
где%
\begin{align}
\mathsf{K}^{\mu\nu}  &  =\varepsilon^{\alpha\beta\rho\sigma}(\delta_{0}^{\mu
}\delta_{\alpha}^{\nu}{}^{\left(  0\right)  }\mathsf{g}_{1\beta}{}%
^{(0)}\mathsf{g}_{2\rho}{}^{(0)}\mathsf{g}_{3\sigma}+\delta_{1}^{\mu}%
\delta_{\beta}^{\nu}{}^{\left(  0\right)  }\mathsf{g}_{0\alpha}{}%
^{(0)}\mathsf{g}_{2\rho}{}^{(0)}\mathsf{g}_{3\sigma}+\nonumber\\
&  +\delta_{2}^{\mu}\delta_{\rho}^{\nu}{}^{(0)}\mathsf{g}_{0\alpha}{}^{\left(
0\right)  }\mathsf{g}_{1\beta}{}^{(0)}\mathsf{g}_{3\sigma}+\delta_{3}^{\mu
}\delta_{\sigma}^{\nu}{}^{(0)}\mathsf{g}_{0\alpha}{}^{\left(  0\right)
}\mathsf{g}_{1\beta}{}^{(0)}\mathsf{g}_{2\rho}),
\end{align}%
\begin{align}
F^{\mu\nu\alpha\beta}  &  =\varepsilon^{\chi\omega\rho\sigma}(\delta_{0}^{\mu
}\delta_{\chi}^{\nu}\delta_{1}^{\alpha}\delta_{\omega}^{\beta}{}%
^{(0)}\mathsf{g}_{2\rho}{}^{(0)}\mathsf{g}_{3\sigma}+\delta_{0}^{\mu}%
\delta_{\chi}^{\nu}\delta_{2}^{\alpha}\delta_{\rho}^{\beta}{}^{\left(
0\right)  }\mathsf{g}_{1\omega}{}^{(0)}\mathsf{g}_{3\sigma}+\delta_{0}^{\mu
}\delta_{\chi}^{\nu}\delta_{3}^{\alpha}\delta_{\sigma}^{\beta}{}^{\left(
0\right)  }\mathsf{g}_{1\omega}{}^{(0)}\mathsf{g}_{2\rho}+\nonumber\\
&  +\delta_{1}^{\mu}\delta_{\omega}^{\nu}\delta_{2}^{\alpha}\delta_{\rho
}^{\beta}{}^{\left(  0\right)  }\mathsf{g}_{0\chi}{}^{(0)}\mathsf{g}_{3\sigma
}++\delta_{1}^{\mu}\delta_{\omega}^{\nu}\delta_{3}^{\alpha}\delta_{\sigma
}^{\beta}{}^{(0)}\mathsf{g}_{0\chi}{}^{(0)}\mathsf{g}_{2\rho}+\delta_{2}^{\mu
}\delta_{\rho}^{\nu}\delta_{3}^{\alpha}\delta_{\sigma}^{\beta}{}%
^{(0)}\mathsf{g}_{0\chi}{}^{(0)}\mathsf{g}_{1\omega}).
\end{align}

Общее
выражение
для
разложения
$\sqrt{g}$ принимает
вид
\begin{equation}
\sqrt{g}=\sqrt{{}^{\left(  0\right)  }g}-\frac{\mathsf{h}_{\mu\nu}%
\mathsf{K}^{\mu\nu}\left(  ^{\left(  0\right)  }\mathsf{g}\right)
+\mathsf{h}_{\mu\nu}\mathsf{h}_{\alpha\beta}\mathsf{F}^{\mu\nu\alpha\beta
}\left(  ^{\left(  0\right)  }\mathsf{g}\right)  }{2\sqrt{{}^{\left(
0\right)  }g}}-\frac{\left(  \mathsf{h}_{\mu\nu}\mathsf{K}^{\mu\nu}\left(
^{\left(  0\right)  }\mathsf{g}\right)  \right)  ^{2}}{8\sqrt{{}\left(
^{\left(  0\right)  }g\right)  ^{3}}}, \label{sqrt_g}%
\end{equation}
где $^{\left(  0\right)  }g=\left\vert \det\,^{(0)}%
\mathsf{g}_{\mu\nu}\right\vert .$

В статье
рассматривается стандартный 
случай (разложение над плоской метрикой) $^{(0)}\mathsf{g}_{\mu\nu}%
=\eta_{\mu\nu}$, тогда $^{\left(  0\right)  }g,$
$K^{\mu\nu},$ $F^{\mu\nu\alpha\beta}$
переходят в
\begin{equation}
\det\left(  ^{(0)}\mathsf{g}_{\mu\nu}\right)  ={}^{\left(  0\right)
}g=-1,\ \mathsf{K}^{\mu\nu}=-\eta^{\mu\nu},\ \ \mathsf{F}^{\mu\nu\alpha\beta
}=\frac{1}{2}\left(  \eta^{\alpha\mu}\eta^{\beta\nu}-\eta^{\mu\nu}\eta
^{\alpha\beta}\right)
\end{equation}
и мы имеем
для (\ref{sqrt_g})%
\begin{equation}
\sqrt{g}=1-\frac{-h+\mathsf{h}_{\mu\nu}\mathsf{h}_{\alpha\beta}\frac{1}%
{2}\left(  \eta^{\alpha\mu}\eta^{\beta\nu}-\eta^{\mu\nu}\eta^{\alpha\beta
}\right)  }{2}-\frac{\left(  -h\right)  ^{2}}{8}=1+\frac{1}{2}h-\frac{1}%
{4}\mathsf{h}_{\mu\nu}\mathsf{h}^{\mu\nu}+\frac{h^{2}}{8}.
\end{equation}

Важным
является тот
факт, что это
выражение
совпало с (\ref{sqrt}).

\end{document}